
\input harvmac


\hyphenation{anom-aly anom-alies coun-ter-term coun-ter-terms
dif-feo-mor-phism dif-fer-en-tial super-dif-fer-en-tial dif-fer-en-tials
super-dif-fer-en-tials reparam-etrize param-etrize reparam-etriza-tion}


%
%
\newwrite\tocfile\global\newcount\tocno\global\tocno=1
\ifx\bigans\answ \def\tocline#1{\hbox to 320pt{\hbox to 45pt{}#1}}
\else\def\tocline#1{\line{#1}}\fi
\def\toclead{\leaders\hbox to 1em{\hss.\hss}\hfill}
\def\tnewsec#1#2{\xdef #1{\the\secno}\newsec{#2}
\ifnum\tocno=1\immediate\openout\tocfile=toc.tmp\fi\global\advance\tocno
by1
{\let\the=0\edef\next{\write\tocfile{\medskip\tocline{\secsym\ #2\toclead\the
\count0}\smallskip}}\next}
}
\def\tnewsubsec#1#2{\xdef #1{\the\secno.\the\subsecno}\subsec{#2}
\ifnum\tocno=1\immediate\openout\tocfile=toc.tmp\fi\global\advance\tocno
by1
{\let\the=0\edef\next{\write\tocfile{\tocline{ \ \secsym\subsecsym\
#2\toclead\the\count0}}}\next}
}
\def\tappendix#1#2#3{\xdef #1{#2.}\appendix{#2}{#3}
\ifnum\tocno=1\immediate\openout\tocfile=toc.tmp\fi\global\advance\tocno
by1
{\let\the=0\edef\next{\write\tocfile{\tocline{ \ #2.
#3\toclead\the\count0}}}\next}
}
%
%

%
%
%
%

%
\long\def\optional#1{}
\def\cmp#1{#1}         
%
%
\let\narrowequiv=\equiv
\def\equiv{\;\narrowequiv\;}
\let\narrowtilde=\tilde
\def\tilde{\widetilde}
\fontdimen16\tensy=2.7pt\fontdimen17\tensy=2.7pt 



%

\def\la{\lambda}

\def\al{{\alpha}}

%
%

\def\CT{{\cal T}}

\def\CO{{\cal O}}
\def\CM{{\cal M}}\def\CCM{$\CM$}
\def\CMH{\widehat\CM}\def\CCMH{$\CMH$}

\def\CE{{\cal E}}\def\CCE{$\CE$}

\def\CN{{\cal N}}
\def\CD{{\cal D}}

%
%
%
\def\bivector#1{{\buildrel \leftrightarrow\over #1}}
%

\def\lfr#1#2{{\textstyle{#1\over#2}}} 



\def\splitexact#1#2{\mathrel{\mathop{\null{
\lower4pt\hbox{$\rightarrow$}\atop\raise4pt\hbox{$\leftarrow$}}}\limits
^{#1}_{#2}}}

%
%
\def\pa{\partial}

\def\pd#1#2{{\partial #1\over\partial #2}} 
\def\pdspace#1#2{{\partial #1\over\partial \vbox{\vskip.5pt\hbox
 {$\scriptstyle #2$}} }} 
%
%
\def\zb{{\bar{\vphantom\i z}}}  
\def\ab{{\bar{\vphantom\i a}}}  
%
%

\def\dd{\mskip 1.3mu{\rm d}\mskip .7mu} 


\def\det{\hbox{det$\,$}}                
\def\BRST{{\caps brst}}
%
%

\def\IM{isomorphism}
\def\smf{super\-manifold}

\def\RS{Riemann surface}
\def\eg{{\it e.g.}}\def\ie{{\it i.e.}}
\def\ho{holomorphic}

\def\SC{super\-con\-for\-mal}

\def\s{super}

\def\topo{top\-o\-log\-i\-cal}
%
%
\def\cmfontflag{cm}
\ifx\fontflag\cmfontflag
\else
 
\fi



\font\df=cmssbx10                      

\global\newcount\figno \global\figno=1
\newwrite\ffile
\def\pfig#1#2{Fig.~\the\figno\pnfig#1{#2}}
\def\pnfig#1#2{\xdef#1{Fig. \the\figno}%
\ifnum\figno=1\immediate\openout\ffile=figs.tmp\fi%
\immediate\write\ffile{\noexpand\item{\noexpand#1\ }#2}%
\global\advance\figno by1}

%
%
\def\tfig#1{Fig.~\the\figno\xdef#1{Fig.~\the\figno}\global\advance\figno by1}





\def\t#1#2#3{t_{#1#2}^{\phantom{#1#2}#3}}



\def\bz{{\bf z}}                     

%
%

%


\def\inbar{\,\vrule height1.5ex width.4pt depth0pt}
\def\IB{\relax{\rm I\kern-.18em B}}
\def\IC{\relax\hbox{$\inbar\kern-.3em{\rm C}$}}
\def\ID{\relax{\rm I\kern-.18em D}}
\def\IE{\relax{\rm I\kern-.18em E}}
\def\IF{\relax{\rm I\kern-.18em F}}
\def\IG{\relax\hbox{$\inbar\kern-.3em{\rm G}$}}
\def\IH{\relax{\rm I\kern-.18em H}}
\def\II{\relax{\rm I\kern-.18em I}}
\def\IK{\relax{\rm I\kern-.18em K}}
\def\IL{\relax{\rm I\kern-.18em L}}
\def\IM{\relax{\rm I\kern-.18em M}}
\def\IN{\relax{\rm I\kern-.18em N}}
\def\IO{\relax\hbox{$\inbar\kern-.3em{\rm O}$}}
\def\IP{\relax{\rm I\kern-.18em P}}
\def\IQ{\relax\hbox{$\inbar\kern-.3em{\rm Q}$}}
\def\IR{\relax{\rm I\kern-.18em R}}
\font\cmss=cmss10 \font\cmsss=cmss10 at 10truept
\def\IZ{\relax\ifmmode\mathchoice
{\hbox{\cmss Z\kern-.4em Z}}{\hbox{\cmss Z\kern-.4em Z}}
{\lower.9pt\hbox{\cmsss Z\kern-.36em Z}}
{\lower1.2pt\hbox{\cmsss Z\kern-.36em Z}}\else{\cmss Z\kern-.4em Z}\fi}
\def\IGa{\relax\hbox{${\rm I}\kern-.18em\Gamma$}}
\def\IPi{\relax\hbox{${\rm I}\kern-.18em\Pi$}}
\def\ITh{\relax\hbox{$\inbar\kern-.3em\Theta$}}
\def\IOm{\relax\hbox{$\inbar\kern-3.00pt\Omega$}}




\def\spec#1{{\df #1}}


\def\lst{\lambda^s\lambda^t}
\def\epdn#1#2#3{\epsilon_{#1#2#3}}
\def\order#1{\CO(\la^{#1})}

\def\frame#1#2{\hat e^{\{#1\}}_#2}
\def\Frame#1#2{\hat E^{\{#1\}}_#2}
\def\lrm{{\buildrel \leftrightarrow\over M}}
\def\IM{iso\-mor\-phism}
\def\TG{topological gravity}
\def\SR{semi\-rigid}
\def\SC{super\-con\-formal}
\def\ct{coordinate trans\-formation}
\def\ntilde{\narrowtilde}
\font\caps=cmcsc10
\Title{\vbox{\hbox{UPR--0477T}}}
{Semirigid Geometry}

\centerline{Suresh Govindarajan, Philip Nelson and Eugene Wong}\smallskip
\centerline{Physics Department}
\centerline{University of Pennsylvania}
\centerline{Philadelphia, PA 19104 USA}
\bigskip
\bigskip

We provide an intrinsic description of $N$-super \RS s and $TN$-\SR\
surfaces. Semirigid surfaces occur naturally
in the description of topological gravity  as well as topological
supergravity. We show that such surfaces are obtained by an
integrable reduction of the structure group of a complex supermanifold.
We also discuss the \s moduli spaces of $TN$-\SR\ surfaces and their
relation to the moduli spaces of $N$-\s\ \RS s.
\vskip1cm

\Date{9/91}\noblackbox                 %
\lref\JDTG{J. Distler, ``2-d quantum gravity, topological field theory,
and the multicritical matrix models,'' Nucl. Phys. {\bf B342} (1990) 523.}%
\lref\DDK{P. Di\thinspace Francesco, J. Distler, and D. Kutasov,
``Superdiscrete series coupled to 2d supergravity,'' Mod. Phys. Lett.
{\bf A5} (1990) 2135.}%
\lref\EWTQFT{E. Witten, ``Topological quantum field theory,'' Commun.
Math. Phys. {\bf 117} (1988) 353.}
\lref\EWTG{E. Witten, ``On
the structure of the topological phase of two-dimensional gravity,''
Nucl. Phys. {\bf B340} (1990) 281.}%
\lref\LPWTG{J. Labastida, M. Pernici, and E. Witten, ``Topological
gravity in two dimensions,'' Nucl. Phys. {\bf B310} (1988) 611.}%
\lref\MSTG{D. Montano and J. Sonnenschein, ``Topological strings,''
Nucl. Phys. {\bf B313} (1989) 258; ``The topology of moduli space and quantum
field theory,'' Nucl. Phys. {\bf B324} (1989) 348.}%
\lref\MPTG{R. Myers and V. Periwal, Nucl. Phys. {\bf B333} (1990) 536.}%
\lref\OSvB{S. Ouvry, R. Stora, and P. van Baal, Phys. Lett. {\bf220B}
(1989) 159.}%
\lref\BSTG{L. Baulieu and I. Singer, ``Conformally invariant gauge
fixed actions for 2D topological gravity,'' Commun. Math. Phys.
{\bf135} (1991) 253.}%
\lref\SBGTG{S. Giddings, ``Cohomological field theory from field-space
cohomology,'' preprint UCSBTH-90-63 (1990).}
\lref\VVTG{E. Verlinde and H. Verlinde, ``A solution of two-dimensional
topological gravity,'' Nucl. Phys. {\bf B348} (1991) 457.}%
\lref\EWTSM{E. Witten, ``Topological sigma models,'' Commun. Math. Phys.
{\bf118} (1988) 411.}%
\lref\EgYa{T. Eguchi and S.-K. Yang, ``N=2 superconformal models as
topological field theories,'' Mod. Phys. Lett. {\bf A5} (1990) 1693.}%
\lref\Keke{K. Li, ``Topological gravity with minimal matter,'' preprint
CALT-68-1662 (1990); ``Recursion relations in topological gravity with
minimal matter,'' preprint CALT-68-1670 (1990).}%
\lref\VafaTLG{C. Vafa, ``Topological Landau-Ginzburg models,'' Mod.
Phys. Lett. {\bf A6} (1990) 337.}%
\lref\DVVTS{R. Dijkgraaf, H. Verlinde, and E. Verlinde,
``Topological strings
in $D < 1$,'' preprint PUPT-1204 (1990), Nucl. Phys. B, in press.}
\lref\Ntwo{
M. Ademollo {\it et al.}, Phys. Lett. {\bf 62B} (1976) 105;
Nucl. Phys. {\bf B111} (1976) 77.}
\lref\JDCNtwo{
J. Cohn, ``N=2 super \RS s,'' Nucl. Phys. {\bf B284}
(1987) 349.}
\lref\Horne{
J. Horne, ``Superspace versions of topological theories,''
Nucl. Phys. {\bf B318} (1989) 22.}%
\lref\Mancrit{
Yu.~Manin, {``Critical dimensions of string
theories,''} Funk.\ Anal.\ {\bf20} (1986) 88 [=Funct. Anal. Appl. {\bf20}
(1987) 244].}
\lref\FrSB{D. Friedan in M. Green and D. Gross, eds. {\sl Unified string
theories} (World Scientific, 1986).}%
\lref\LSMF{
P. Nelson, ``Lectures on supermanifolds and strings," in
{\sl Particles, strings, and supernovae,} ed. A. Jevicki and C.-I. Tan
(World Scientific, 1989).}
\lref\Cohnb{
J. Cohn, ``Modular geometry of
superconformal field theory," Nucl. Phys. {\bf B306} (1988) 239.}
\lref\CIGVO{
P. Nelson, ``Covariant insertions of general vertex operators,"
Phys. Rev. Lett. {\bf62} (1989) 993.}%
\lref\TCCT{J. Distler and P. Nelson, ``Topological couplings and
contact terms in 2d field theory,''
Commun. Math. Phys. {\bf 138} (1991) 273.}%
\lref\DRS{S. Dolgikh, A. Rosly, and A. Schwarz, ``Supermoduli spaces,''
Commun. Math. Phys. {\bf135} (1990) 91.}%
\lref\semirig{J. Distler and P. Nelson, ``Semirigid supergravity,''
Phys. Rev. Lett. {\bf66} (1991) 1955.}
\lref\DVVCarg{R. Dijkgraaf, H. Verlinde, and E. Verlinde, ``Notes on
topological string theory and 2-D quantum gravity,'' PUPT-1217.}%
\lref\EWTGlect{E. Witten, ``Two dimensional gravity and intersection
theory on moduli space,'' preprint IASSNS-HEP-90/45.}%
\lref\ChKua{D. Chang and A. Kumar, ``\BRST  quantization of \SC\
theories,'' Phys. Rev. {\bf D35} (1987) 1388.}
\lref\MAS{D. Montano, K. Aoki, and J. Sonnenschein, ``Topological \s
gravity in two dimensions,''  Phys. Lett. {\bf B247} (1990) 64.}
\lref\HuLi{J. Hughes and K. Li, ``Free-field formulation of
topological \s
gravity,'' CALT--68--1695.}
\lref\CoCo{S. Cordes and M. O'Connor, ``Topological \s gravity,'' NUB--3017.}
\lref\ScSe{A. Schwimmer and N. Seiberg, ``Comments on the $N=2,3,4$ \SC\
algebras in two dimensions,'' Phys. Lett. {\bf B184} (1987) 191.}
\lref\ChKub{D. Chang and A. Kumar, ``Representations of $N=3$ \SC\
algebra,'' Phys. Lett. {\bf B193} (1987) 181.}
\lref\topsug{S. Govindarajan, P. Nelson and S-J. Rey,
``Semirigid construction
of topological supergravities,''  UPR0472-T = UFIFT-HEP-91-10
(1991), Nucl. Phys. {\bf B}, in press.}
\lref\QiuRey {Z. Qiu and S.-J. Rey, `` Topological N=1 superstrings
in $ \hat c < \infty$,''
 Univ. of Florida preprint, UFIFT-HEP-91-11 (1991).}
\lref\Schou {K. Schoutens, `` $O(N)$-extended superconformal field theory
in superspace, Nucl. Phys. {\bf B295} [FS21] (1988) 634.}
\lref\dil {J. Distler and P. Nelson, ``The dilaton equation in
semirigid string theory,''  PUPT-1232 = UPR0428T (1991), Nucl. Phys.
{\bf B}, in press.}
\lref\punc { E. Wong, to appear.}
\lref\sbgpn {S. B. Giddings and P. Nelson,
`` The Geometry of super \RS s,''
Commun. Math. Phys. {\bf 116}, (1988) 607.}
\lref\bfs{ M. A. Baranov, I. V. Frolov and A.S. Schwarz, ``Geometry of
two-dimensional superconformal field theories,'' Teor. Mat. Fiz, {\bf
70}(1987) = Theor. Math. Phys. {\bf 70}(1987) 92.}
\lref\ssca{S. S. Chern, ``The geometry of $G$-structures,'' Bull. Am.
Math. Soc. {\bf 72}(1966)  167.}
\lref\sscb{S. S. Chern,   {\sl Complex manifolds without potential
theory}, (Springer-Verlag, 1979), second edition.}
\lref\vginteg{V. Guillemin,``The Integrability problem for
$G$-Structures,''Trans. Am. Math. Soc. {\bf 116}(1965) 544.}
\lref\nninteg{A. Newlander and L. Nirenberg,``Complex analytic
coordinates in almost complex manifolds,'' Ann. of Math.,
{\bf 65}(1957) 391.}
\lref\nakh{M. Nakahara, {\sl Geometry, topology and physics},
(Adam Hilger, 1990).}
\lref\stern{S. Sternberg, {\sl Lectures on differential geometry,}
 (Chelsea Publishing Company, 1983), second edition.}
\lref\Mancrit{Yu.~Manin, \cmp{``Critical dimensions of string
theories,''} Funk.\ Anal.\ {\bf20} (1986) 88 [=Funct. Anal. Appl. {\bf20}
(1987) 244].}%
\lref\rsw{S. Weinberg, {\sl Gravitation and cosmology},
(Wiley, New York, 1972).}

\newsec{Introduction}
Semirigid surfaces
have been shown~\semirig\topsug\
to provide a geometric framework to describe $2d$ topological gravity
and \s gravity.
For example, in the simplest theory the dilaton as well as the puncture
equations
have been proven using the semirigid formalism \dil\punc. In this paper,
we provide an intrinsic or coordinate invariant
definition of \SR\ super Riemann surfaces ($SSRS$) as well as
ordinary super \RS s ($SRS$). The discussion of $SRS$  is a
natural extension to similar discussions provided in~%
\ref\Schwarzsmf{A.S. Schwarz, ``Supergravity, complex geometry, and
$G$-structures,'' Commun. Math. Phys. {\bf 87} (1982) 37.}%
\ and applied in~%
\sbgpn\ for the case of $N=1~SRS$ and in~\ref\Lott{J. Lott, ``Torsion
constraints in supergeometry,'' Commun. Math. Phys. {\bf133} (1990) 563.}%
\ for $N=2$; the framework
follows Cartan's theory of $G$-structures.
(For an introduction to $G$-structures, see for example
\stern\ssca\sbgpn\LSMF.)
We show that these structures subject to some conditions
called ``torsion constraints'' are
integrable, which relates our intrinsic definition to the coordinate
dependent definitions.

We will first discuss the various definitions and
illustrate $G$-structures via two examples in sect.~2. We also find the
appropriate group $G$ for \SC\ and \SR\ surfaces and the corresponding
torsion constraints. Sect.~3
 deals with showing that the $G$-structures we impose are
integrable provided the constraints are satisfied.
Briefly the results are as follows. If we begin with a complex \smf,
then $N$-$SRS$ have no essential torsion constraints, generalizing
Baranov, Frolov, and Schwarz~\bfs, who considered
$N=1$.\foot{This generalization was asserted in the appendix to \DRS.
The constraints found in~%
\ref\How{P. Howe, \cmp{``Super Weyl transformations in two dimensions,"}
J. Phys. {\bf A12} (1979) 393.}%
\ and discussed in \sbgpn\ arise when we begin with a {\it real\/}
\smf.}\  We will refer to \SR\ surfaces with $N$-supersymmetry as
``topological $N$-$SRS$,'' or $TN$ for short.
$TN=0$ surfaces have a rather trivial essential
constraint while $TN=1$ surfaces have several. Both in the usual and
in the topological case
the category of surfaces with appropriate $G$-structures, integrable in
the sense we will specify, is equivalent to the corresponding category
of surfaces with appropriate patching data. (Actually we will limit
ourselves to proving this for $N\le 3$ and $TN\le 1$ to keep the
algebra simple.) In particular there are no
second-order conditions for flatness, just as for ordinary $N=0$
conformal structures. Throughout this paper we will consider only
untwisted \SC\ and \SR\ structures, since our focus is primarily on
local properties. The integrability results we prove will also apply
to the study of twisted surfaces.

We should comment on the relation of this work to \semirig\topsug. In
these papers the coordinate definition of \SR\ surfaces was used. The
interpretation of such surfaces as having a special $G$-structure was
crucial for finding the right patching maps, but no attempt was made
to prove the equivalence of the two approaches, \ie\ the theorem that
every integrable $G$-structure gave a \SR\ surface. That is what we do
here.

\newsec{$G$-structures on manifolds and supermanifolds}
We begin by stating the problem, then recall the general idea of
$G$-structures with some examples.
\subsec{Patch definition of SRS and SSRS}
One way of defining $SRS$ or $SSRS$ is to cut a \s
manifold into patches, put coordinates on them and sew them back
together with transition functions given by superconformal or
semirigid coordinate transformation.
Let us begin with $SRS$.
Generalizing the $N=1$ superconformal
transformation~\FrSB\bfs\JDCNtwo\DRS, we start with  $\spec C ^{1|N}$
and define for $i=1, \ldots, N$
\eqn\eDi{D_i={\pa\over\pa\theta^i} +
g_{ij}\theta^j\;{\pa\over\pa z}
}
where $g_{ij}=\delta_{ij}$.
We impose the condition that
$\{D_i\}$ transform linearly among
themselves (not mix with $\pa\over\pa z$) under a \SC\ \ct\
$(z, \theta^i) \rightarrow ({\tilde z}, {\tilde \theta^i})$.
This condition resembles the one for a complex manifold, where the good \ct s
do not mix the $\pa_{z^i}$ with the $\pa_{{\zb }^i}$.
Thus,
\eqn\eMtm{
D_i=F_i^{\ j}{\tilde D}_j\quad;\qquad
F_i^{\ j}=D_i{\tilde \theta}^{\ j}\quad,
}
where ${\tilde D_i}=\pdspace{}{\narrowtilde\theta^i}
+g_{ij}\narrowtilde\theta^j{\pa\over\pa\ntilde z}$ and $F$ is some
invertible matrix of functions.
It follows that the \SC\ transformations are those for which
\eqn\escc{
D_i {\tilde z} = g_{jk} {\tilde\theta^j}D_i {\tilde\theta^k}.
}
An $N$-\SC\ surface is then just a \smf\ patched together from pieces of
$\spec{C}^{1| N}$ related by $N$-\SC\ transition functions.

Semirigid surfaces (or SSRS) are  patched together by restricted
\SC\ transition functions.
The restriction imposed is that $\theta^+$ be global, where $\theta^\pm
\equiv {1 \over {\sqrt 2}}(\theta^1 \pm i \theta^2)$.
For instance,
to obtain the $TN=0$ semirigid \ct s,
we start with $N=2$ \SC\ \ct s and
impose ${\tilde \theta^+}= \theta^+$.  This restriction together with
\escc\  fixes the \ct s on
the rest of the coordinates.  Such restricted \ct s then provide the transition
functions to build a $TN=0$ $SSRS$ \semirig.
One can similarly obtain $TN=1$ \SR\ \ct s from $N=3$ \SC\ \ct s
by the same method.

Although this method of deriving $SRS$ and $SSRS$ is adequate for doing
physics, there are at least two features that are buried in them.
One would like to classify the \SC\ or \SR\ \ct s as
being  \ct s which preserve some geometrical object.  This object
is not obvious using the above patch construction.
In addition, to find the \SC\ or \SR\ moduli space, one would like to
have a coordinate invariant definition of $SRS$ or $SSRS$ so that it
is clear that deformations of their structure are not artifacts of \ct
s. This is of interest when one studies the moduli space of these
surfaces, where one's interest is to find deformations which cannot be
undone by allowed coordinate transformations.

We will provide such an invariant
description in the sequel by means of $G$-structures.  To prove that the
patch definition is equivalent to the intrinsic definition (\ie\
the one using
$G$-structures), we will show that a manifold constructed by the above
 patching
functions implies a $G$-structure.  To invert this correspondence and so
establish equivalence we will ask whether {\it every} $G$-structure
arises by this construction. In
general this last step requires that the given
$G$-structure be ``integrable,'' a concept whose meaning we will recall in the
following examples. We will find the appropriate integrability
conditions in sect.~2.3 and show that they really do lead to an
equivalence between the patch and $G$-structure definitions. While this
is not too difficult for $TN=0$, it does require some work for $TN=1$,
\ie\ for \topo\ \s gravity.

\subsec{Two examples}
In this subsection, we will illustrate $G$-structures and  the
question of their integrability~\ssca\vginteg. We will also
demonstrate how one obtains  \ct s which preserve the $G$-structure
chosen. This enables us to relate this definition to the patch
definition once integrability is proved.

Suppose we are given a smooth manifold.  Then its tangent space can be
locally spanned by a field of frames $\{e_a\}$.
However, there are in general no global frames.
In order to obtain a global structure, we define
an equivalence {\it class} of frames.  The equivalence relation is given by
a group $G$ of matrices  whose elements act on the frames, that is,
  $\{K_a^{\ b}\, e_b\}$ is defined to be equivalent
to $\{e_a\}$, where $K$ is a function with values in $ G$.
Without any extra structure beyond smoothness, all we can say about
the matrices ${K_a}^b$ is that they
belong to the group $GL(n,{\spec R})$.
However, with additional structures,
the structure group can be reduced to a subgroup of $GL(n,{\spec R})$.
The structure group can be thought of
as the local symmetry group of a physical theory defined on the
manifold.
In general, not all manifolds admit a
reduction of structure group due to possible global
obstructions~\ssca\foot{We will not consider such obstructions
because they are not relevant in establishing the equivalence between
the patch and intrinsic definitions.}.
Also, there are  geometrical structures like connections and
projective structures that are not $G$-structures.
What we will see in this paper is that $SRS$ and $SSRS$ as defined in
sect.~2.1 {\it do} arise as
reductions of the structure group of a \smf.

We first consider a smooth manifold with additional structure provided by
 a metric \eqn\emetri{g=g_{ab} e^a\otimes e^b\quad,}
where $e^a$ is the dual to the frame $e_a$.
Since a metric provides information about the length of a vector, it
selects out from the classes of frames $\{e_a\}$
acted on by elements of the group $GL(n,{\spec R})$
those that are orthonormal, that is, $g_{ab} = \delta_{ab}$.
The structure group that acts on the family of orthonormal frames is the
group $O(n)$ leaving $\delta_{ab}$ invariant.
Thus we have a reduction of structure group from
$GL(n,{\spec R})$ to $O(n)$ imposed by the additional structure,
the metric.
Conversely, given a reduction of structure group to $O(n)$,
it induces a metric on the manifold: we simply
substitute any good frame into \emetri.
Like the metric, the imposition of a $G$-structure on a manifold is
an intrinsic concept.
Note that the more structures one imposes,
the smaller the class of good frames. For example, imposing in
addition an orientation lets us restrict further to the class of {\it
oriented} orthonormal frames; these are related by the smaller group
$SO(n)$.

For our second example consider the case of a $2n$-dimensional
 manifold $M$ endowed with an
almost complex structure, specified by a tensor $J$
similar to the metric.
The tensor is given at a point $P$ by $J_P : T_PM \rightarrow T_PM$
everywhere satisfying  $J_P^2 =-\spec I$.
When diagonalized, $J$ splits the complexified tangent $T_cM$ into holomorphic
(with eigenvalue $i$) and antiholomorphic (with eigenvalue $-i$)
tangent spaces.  We can use $J$ to define good frames $\{e_{a},
e_{\ab}\}$ as those for which $e_a$ are $+i$ eigenvectors and
$e_\ab$ are the complex conjugates of $e_a$,
 $a=1, \ldots, n$. Then
\eqn\edJ{
J=i(e_{a} \otimes e^{a} -
e_{\ab} \otimes e^{\ab}) \quad.}
$J$ thus selects out from the class of frames related by $GL(2n,{\spec R})$
a smaller class related by $GL(n,{\spec C})$, since $J$ is invariant
only under $GL(n,{\spec C})$ transformation of frames.
Conversely, given a reduction of structure group to $GL(n,{\spec C})$,
which gives us the class of good frames $\{e_{a}$,
$e_{\ab}\}$,  we can obtain $J$ by substituting any good frame in~\edJ.
Thus an almost complex structure is nothing but a
 $GL(n,{\spec C})$  structure, %
an equivalence
class of frames $\{e_{a},e_{\bar a} \}$ where any two frames are
related by a complex matrix of the form %
\eqn\eege{
\pmatrix{e'_{a} \cr e'_{\bar a} \cr} =
\pmatrix{A&0 \cr 0&{\bar A}} \pmatrix{ e_{a} \cr e_{\bar a}}.
}
$\bar A$ is the complex conjugate of the invertible matrix $A$.

We have given a coordinate invariant characterization of a $G$-structure.
But sometimes it is convenient to use coordinates. Since a
$G$-structure makes sense even locally, let us first consider the
problem of specifying one on an open set $U$ of $\spec R^n$. For any
choice of coordinates $\{x^a\}$ on $U$ we first choose a
standard frame given by some universal rule. For example in Riemannian
geometry we choose
$\hat e_a^{\{x\}} = {\pa \over {\pa x^a}}$.
(We will choose a more complicated standard frame in the \SC\
 and \SR\ cases.)
If we begin with a different set of coordinates $\{y^a\}$, in general
the two frames $\hat e_a^{\{x\}}$, $\hat e_a^{\{y\}}$ do not agree.
However if we  arrange for them to agree {\it modulo} a
$G$-transformation then they do define the same $G$-structure.
This happens when
\eqn\econdn{\frame {y}a|_P=K(P)_a^{\ b}\frame {x}b|_P\quad}
for some function $K$ in $G$. Since $G$ is a group, the set of all
coordinate transformations $y(x)$ defined by \econdn\ is a
group too; we call it the group of $G$-coordinate transformations, or
simply the ``good'' transformations.

Thus one way to specify a $G$-structure on a manifold $M$ is to
present an atlas of coordinate charts $U_\alpha$ with coordinates
$x_\alpha$ all related on patch overlaps by $G$-transformations.

Let us illustrate the above discussion with our two examples. In
Riemannian geometry the only coordinate transformations preserving the
standard frame up to $O(n)$ are the ones preserving the standard
metric, \ie\ the rigid Euclidean motions. For the almost-complex
structure example things are more interesting. Given a choice of real
coordinates $\{u^a,v^a\}$, $a=1,\ldots, n$ we let $z^a=u^a+{\rm i}
v^a$ and take the standard frame to be $\frame {z}a=\pd{}
{z^a},\frame{z}\ab=\pd{}{\zb^a}$.
Let $\{w^a, {{\bar w}^a}\} $ be another complex
local coordinate with standard frame $\{\pa_{w^a}, \pa_{{\bar w}^a}\}$.
On the overlap, let $w$ and $z$ be related
by a \ct\ $w^a=w^a(z^b,{{\zb }^b})$ so that
\eqn\eegz{
 \pmatrix{ \pa_{z^a} \cr \pa_{{\zb }^a}}
= M \pmatrix{\pa_{w^a}\cr \pa_{{\bar w}^a}},\quad
  {\rm where}\quad
M = \pmatrix{\pa_{z^a} w^b & \pa_{z^a} {{\bar w}^b} \cr
         \pa_{{\zb }^a} w^b & \pa_{{\zb }^a} {{\bar w}^b} \cr }.
}
For $w$ and $z$ to be complex coordinates for the same complex
structure, we need $M$ to be of the form \eege.
This means that the ``good'' \ct s preserving the complex structure
 are  {\it holomorphic} maps.

More generally, a manifold obtained by patching together coordinate charts
by a class of $G$-transformations gets a $G$-structure. Clearly if we
replace each local coordinate $x_\alpha^a$ by $y^a_\al=\psi_\al(x^b_\alpha)$
where $\psi_\alpha$ is itself a $G$-transformation, we determine exactly
the same $G$-structure.

We would also like to show the converse: a manifold equipped with a
$G$-structure can {\it always} be constructed from a set of
 ``good'' transition functions. In
fact this converse is not always true. To find out when it is so,
we introduce coordinate patches on the manifold with the
$G$-structure.  We seek coordinates $\{x_\alpha\}$ on a local patch
$U_\alpha$ such that the standard
frame $\{\hat e_a^{\{x_\al\}}\}$ determines the given $G$-structure.  Since
 a $G$-structure is given
by an equivalence class of good frames  we are thus seeking
a local coordinate whose standard frame  belongs to the same
equivalence class as the given $\{e_a\}$.  If we can find such a coordinate
system, we then
call the $G$-structure {\it integrable}.
  However, this is in general not
possible unless the frames belonging to the $G$-structure satisfy
certain constraints. After all, $\{x^a\}$ contains only $n=$dim$M$
degrees of freedom, while the given $\{e_a=e_a^\mu\pa_\mu\}$ has $n^2$ minus
the
dimension of $G$. This counting also makes it clear that
different $G$-structures impose different integrability
constraints.  For instance, we will see that the \SC\ structure does
not need any such conditions while the semirigid case needs some first
order constraints. Of course there is more to do than just count
conditions. The statement that a set of local constraints on a
$G$-structure really does suffice to find local coordinates inducing
that structure is called an {\it integrability theorem}.

Let us illustrate these ideas in the two examples given above. For the
case of Riemannian geometry, $G=O(n)$, it turns out that a $G$-structure
is integrable iff its Riemann curvature tensor $R$ vanishes (see for example
\rsw).
That is, if $R\equiv 0$ in the neighborhood of a point, then there exist
local coordinates (called inertial) such that the metric is in the
standard form $g=\delta_{ab} \dd x^a \otimes \dd x^b$.
Comparing this metric with the one specified by the given $O(n)$-structure
 $g=\delta_{ab} e^a \otimes e^b$, we see that the frames are related
by $e_a = {K_a}^b {\pa \over \pa x^b}$, where $K \in O(n)$.
Thus the frame defining the $G$-structure $e_a$ is $G$-equivalent to the
standard frame of some coordinates, which is what we called integrability
earlier.   Notice that the integrability condition is given by
constraining the curvature, a function involving up to second order
derivatives of the original frame. We thus call this
 a second order constraint. %
The condition $R=0$
implies flat space; thus integrability conditions are sometimes called
flatness conditions, even though they may be given by first order
constraints in other cases.

Instead of Riemannian geometry we can enlarge $O(n)$ somewhat to the
group of matrices with $K^{\rm t}g K\propto g$ --- the
conformal group. The obstruction to flatness is now just a part of the
Riemann curvature, namely the Weyl tensor~%
\ref\rHE{S. Hawking and G. Ellis, {\sl The large-scale structure of space
time} (Cambridge, 1973).}%
{}. An important case is two dimensions, where there is no Weyl tensor at
all and {\it every} conformal (or $\spec C^\times$)-structure is integrable.

In the case of an almost complex structure, the counterpart of
the curvature is the Nijenhuis tensor~\nakh, given in terms of $J$ by
\eqn\eNij{
 \CN(X,Y) = [X,Y] + J[JX,Y] + J[X,JY] - [JX,JY].
}
where $X$ and $Y$ are arbitrary vector fields.
The integrability theorem~\nninteg\ says if $\CN\equiv0$, then there exists a
local complex coordinate system $\{z^a\}$ , $i= 1, \ldots, n$ such that
$J$ is of the form \edJ\ with the frames given by $\frame {z}a=\pd{}
{z^a},\frame{z}\ab=\pd{}{\zb^a}$.
Thus $\CN=0$  becomes the flatness condition. It is however a first order
condition unlike the  $O(n)$ case, since \eNij\ clearly involves at
most first derivatives of $J$.
As mentioned above, the ``good'' coordinate transformations (those
preserving $J$) are the \ho\ maps.

Given an {\it integrable} $G$-structure on $M$, we can now return to the
question of whether it can be constructed via patching maps. On each
 coordinate patch choose a coordinate inducing the given
$G$-structure. Then on patch overlaps the chosen coordinates
are related by what we have called a ``good'' or
$G$-transformation: $x_\beta=\phi_{\alpha\beta}(x_\alpha)$.
Hence we can construct $M$ with its $G$-structure from
 patching coordinate charts with the ``good'' coordinate
transformations. Of course on each patch we have some freedom to
redefine the good coordinate $x^a_\al$ by some $G$-transformation
$y^a_\al=\psi_\al(x^b_\alpha)$. This simply corresponds to replacing the
$\{\phi_{\alpha\beta}\}$ by the equivalent family
$\{\psi_\alpha\circ\psi_{\alpha\beta}\circ\psi\inv_\beta\}$ as discussed above.

To summarize, given $G$ and a choice of standard frames we may define
a $G$-manifold as a collection of patching $G$-transformations modulo the
substitution $\{\phi_{\alpha\beta}\}\mapsto\{\psi_\alpha\circ\phi_{\alpha\beta}
\circ\psi\inv_\beta\}$, where $\psi_\alpha$ are themselves
$G$-transformations. Or we may define a $G$-manifold as a smooth manifold
with a collection of frames defined modulo $G$
satisfying appropriate integrability conditions. We have
seen that {\sl these two definitions are
equivalent once the appropriate integrability theorem is established.}

For the case of specifying the $N\ge 1$ superconformal structure, a
coordinate invariant tensor analogous to
the metric $g$ or the tensor $J$ is not known.
However, one can still choose a group $G$ and specify a $G$-structure by
giving a frame defined up to transformations by elements of $G$.
Without the analog of $g$ or $J$, we cannot
define a tensor like $R$ or $\CN$ measuring the local obstruction to
integrability.  Thus, one has to find another way to
give the flatness condition for the case of superconformal structures
or else prove that there is no such condition, that is, all
$G$-structures are flat.  The situation is similar for semirigid structures.

Let us once again use the case of
an almost complex structure on a $2n$ dimensional real manifold to
clarify how first-order flatness conditions can come about.
The flatness condition $\CN=0$ can be replaced
by a condition similar to the one used in the Frobenius integrability
theorem,
namely ${t_{a b}}^{\bar c}=0$, %
 where
\eqn\etordef{[e_A, e_B] = {t_{AB}}^C e_C}
and $A$ denotes either $a,\ab$.
In other words, the Lie bracket of the
holomorphic tangent frames stays in the same subspace.
Conditions of this
type are sometimes called ``essential torsion constraints''\sbgpn.

We now recall a general prescription~\sbgpn\ to obtain the  torsion
constraints with the above example in mind and see that they are
necessary conditions for integrability.
In our examples the structure constants $\hat \t abc$ all vanish when we use
the
standard frame $\{\frame xa \}$ in
$[\frame xa, \frame xb] = \hat \t abc
\frame xc$. (More generally they will at least all
be {\it constants} in the cases of interest.) Of course the same may not be
true when we substitute some other equivalent frame $\{e_a\}$ to get
${t_{ab}}^c$.
We obtain an arbitrary representative of the standard $G$-structure
by letting an arbitrary function in %
$G$ act on the standard frame.  Those
${t_{ab}}^c$ that remain equal to $\hat \t abc$ clearly have
the same values in {\it any} good frame. Thus we have found some
conditions on ${t_{ab}}^c$ which follow from the assumption that our
frame is equivalent to some standard frame. These conditions may be
overcomplete; for example some may be related to others by Jacobi
identities.

In other words given a frame we have found some conditions which must
be met if the corresponding $G$-structure is to be integrable.
These ``torsion constraints''
are first order conditions on any frame representing the given
structure since the Lie bracket entering $t$ contains one derivative. If we
find
that they are also {\it sufficient} for flatness, then we have an
integrability theorem with {\it only} first order constraints. This is
the case for $G=GL(n,\spec C)$ since here the torsion constraints amount
to the vanishing of the Nijenhuis tensor;  it will also be true for
\SC\ and \SR\ geometry. (And as we have mentioned, for \SC\ geometry
there will be
no essential torsion constraints at all.) However as we have seen it is
false for Riemannian geometry.
It is sometimes convenient to impose further $G$-invariant
``inessential'' torsion constraints corresponding to normalization
conditions~\sbgpn, as we will recall below.

We will now apply all these ideas to
the cases of $N$ \SC\ and $TN$ \SR\ structure.

\subsec{Intrinsic Definitions of SRS and SSRS}
We now
provide an intrinsic definition of  $N$ \SC\  structures~\DRS\
generalizing~\bfs\Mancrit. Below we will propose a similar intrinsic
definition of \SR\ structures.
Let $\hat{M}$ be a complex supermanifold of dimension $1|N$
equipped with a holomorphic distribution (subbundle of $TM$)
${\cal E}$ of dimension $0|N$. Given $(\hat M, \CE)$, one can always
define a  symmetric bilinear
$B : {\cal E} \otimes {\cal E} \rightarrow {\cal T}/{\cal E}$,
where ${\cal T}$ is the holomorphic tangent bundle.
The bilinear  is given by
$B(E_i,E_j) \equiv {[E_i,E_j]} ~{{\rm mod} ~\CE}$,
where $[~,~]$ is the graded Lie bracket and $E_i \in {\cal E}$.
Following \bfs\Mancrit\DRS,
we will call $(\hat M,\CE)$ an $N$-$SRS$ if $B$ is non-degenerate.

A $SRS$ can also be regarded as a reduction of the structure group on
$\hat{M}$. We simply declare a frame $\{E_0,\vec E\,\}$ as ``good'' if
$E_0$ is even and $E_i$ are an (odd) frame for the given \CCE.
Then all good frames are
related to one another by elements of a supergroup as follows:
\eqn\eFtm{
\pmatrix{
E_0' \cr\vec E'\cr}=
\left(
\vbox{\offinterlineskip\parindent=0pt
\tabskip=1em plus 1em minus0.5em
\halign{\hfill$#$\hfill\ \vrule&\hfill$#$\hfill\cr
a^2&\vec\omega\cr
&\phantom{\vrule height3pt}\cr
\noalign{\hrule}
&\phantom{\vrule height3pt}\cr
\vec 0&a\lrm\cr
\noalign{\vskip -0.1truein}
}}
\right)
\pmatrix{
E_0 \cr
\vec E\cr}\quad,
}
where $a$ is an invertible even function, $\vec\omega$ are odd
functions, and $\lrm$ is an invertible matrix of even functions.
In order for the set of frames $\{E_i'\}$  to span the same
distribution $\cal E$ as $\{E_i\}$,
we have required the column $\vec 0$.

We can always put a $SRS$ in a more canonical form.
The non-degeneracy condition above implies that the bilinear $B$ is
diagonalizable. Thus we can always use a transformation of the form
\eFtm\ to get from a frame $\{E_0,\vec E\,\}$ to a normalized frame with
\eqn\enondeg{
[{E_i,E_j]} = 2 g_{ij} E_0 ~~{{\rm mod} ~\CE}
}
where $g_{ij}=\delta_{ij}$.
Such normalized good frames are then all related by a smaller group
than \eFtm, in which $M$ is in the orthogonal group $O(N,\spec C)$.
This residual group we will call $G^{N}$, and we will call a
$G^N$-structure an almost \SC\ structure.
Since we can always pass to normalized frames, and the new frame is unique
modulo the residual group, we find that {\sl an $N$-$SRS$ in the above
sense is precisely a reduction of the structure group of $\hat M$ to $G^{N
}$.}
We will prove in sect.~3
that this reduction
$N\leq3$ is always integrable.

We would like to point out that $E_+$ and $E_-$(in a complex basis)
in the $N=2$ case are preserved up to
a multiplicative factor on a $SRS$ because in this basis matrices
in $O(2,{\spec C})$ are diagonal.
Hence the distribution $\cal E$ is split into two line
bundles. This is not true for $N\ge3$, a fact related to the existence
of a nonabelian current algebra in the \SC\ algebra starting at $N=3$.

What are the ``good'' coordinate transformations for this \SC\ structure?
To answer this, and to make precise what we wish to prove in the
integrability theorem, we must specify the standard frames associated
to a coordinate patch. We choose $\Frame\bz0=\pd{}z$, $\Frame\bz
i=D_i$ where $\bz\equiv(z,\vec\theta\,)$ and $D_i$ are defined in \eDi.
We can then identify the $N$-\SC\ \ct s as those complex
coordinate transformations that leave
this structure unchanged along the lines similar to the discussion
below \eegz.
Then the ``good'' coordinate transformations preserving the standard $G$
structure will
take $\bz$ to ${\tilde {\bz}}$ with
\eqn\ectm{
\pmatrix{
\pa_{\ntilde z}\cr\vec {\tilde D}\cr}=
\left(
\vbox{\offinterlineskip\parindent=0pt
\tabskip=1em plus 1em minus0.5em
\halign{\hfill$#$\hfill\ \vrule&\hfill$#$\hfill\cr
a^2&\vec\omega\cr
&\phantom{\vrule height3pt}\cr
\noalign{\hrule}
&\phantom{\vrule height3pt}\cr
\vec 0&a\lrm\cr
\noalign{\vskip -0.1truein}
}}
\right)
\pmatrix{
\pa_z\cr
\vec D\cr}\quad.
}
The set of coordinate transformations in the form of
\ectm\ are given by the
$N$-\SC\ transformations defined by  \eMtm--\escc.
As in the general analysis above, this  leads to a patch definition of super
\RS s.
Once the integrability theorem is proved in sect.~3 we thus have that
every $N$-$SRS$ in the
above sense is also a $SRS$ in the sense of sect.~2.1. %

Next we turn to the \SR\ case.
An almost $TN$-structure is obtained by reduction of
the structure group from an ($N+2$)-\SC\ structure.
Consider the set of frames spanning $\cal E$,
$\{E_i\}=\{E_+,E_r,E_-\}$ where $E_{\pm} = {1\over {\sqrt 2}}
(E_1 \pm i E_2)$ and $r=3,\ldots,N+2$.
The metric $g_{ij}$ in this frame is
\eqn\eoffbasis{
g_{ij}= \pmatrix{0 & \vec 0 & 1\cr
                     \vec 0 & \tilde{g}_{rs} &\vec 0\cr
                        1 & \vec 0 & 0}\quad,
}
where $\tilde{g}_{rs}= \delta_{rs}$.
The reduction from \eFtm\ is specified by the $G$-structure where now
the group consists of matrices of the form
\eqn\essr{%
K=
\pmatrix{
a^2   &\omega_-&\vec\omega &\omega_+\cr
0     & 1 &\vec Y  &  -\half Y \tilde g Y^t   \cr
\vec 0& \vec 0  &a\lrm&  -aM \tilde g Y^t\cr
0 & 0& \vec 0 &a^2}
\quad.}
Here $\vec Y$, $\bivector M$ have %
even elements,
$M \tilde g M^t =  \tilde g $,
$a$ is invertible and the $\omega$ are odd functions. It can
be verified that matrices of type \essr\ form a supergroup $G^{TN}$,
which is a subgroup of $G^{N+2}$. In fact this
 structure group arises by a reduction
from $G^{N+2}$ by imposing extra structure:
we have chosen a $1d$ subbundle $\CD_-\subset\CE\subset\CT$. $\CD_-$
is not trivial; indeed we also choose a (parity-reversing) isomorphism
$\CD_-\cong \CT/\CE$. The good frames are those good \SC\ frames for
which $E_-$ spans $\CD_-$ and corresponds to $E_0$~mod~$\CE$ under the
chosen isomorphism. These frames are then all related by \essr.

The motivation for this
construction is simple for $TN=0$. Any kind of \topo\ field theory should have
a
\s space formulation involving a global, spinless odd coordinate for
bookkeeping. For us this coordinate will be $\theta^+$. For $TN=0$
\essr\ says that ``good'' coordinate transformations take $D_+$ to itself,
and hence they also take $\theta^+$ to itself as desired. For $N>0$
this may not be so clear, but in fact \essr\ again ensures that
the ``good'' $TN$-coordinate transformations are just $N$-\SC\
transformations which keep $\theta^+$ fixed~\topsug.
Note that the $N$-\SC\ structure group is embedded in
that of $TN$ semirigid geometry, $G^{N} \subset G^{TN} \subset
G^{N+2}$
by comparing with \eFtm. This is seen by
setting $\vec Y= \omega_+ =\omega_-=0$. This is why the
$TN$-coordinate transformations include the $N$-\SC\ group and give
rise to \topo\ \s gravity.

In sect.~4 we will find first order constraints which
are sufficient flatness conditions
for the existence of a coordinate system with the standard frames
$G^{TN}$-equivalent to the frames $E_a$ defining the semirigid structure.
Hence as in our general discussion a complex \s
manifold with an integrable $G^{TN}$-structure
is glued together by semirigid transition
functions, which recovers
the patch definition of semirigid surfaces given in sect.~2.1.

\newsec{Superconformal Integrability}
\def\mat#1#2#3{#1_{#2}^{\phantom{#2}#3}}
\def\ord#1{{\cal O}(#1)}
\def\ordl{\ord\la}
In sect.~2.3 we defined an almost superconformal structure.
We shall prove that this reduction is always integrable for $N= 3$;
there are no flatness
conditions to impose in this case. The cases $N<3$ are much easier
and can easily be obtained from our derivation. We expect $N=4$ to be
similar.

We are given a distribution \CCE\ which satisfies the
non-degeneracy condition. As we have discussed above we can always
choose a frame $\{E_0,\vec E\,\}$ with $\vec E$ spanning \CCE\ and
satisfying \enondeg, or in the notation of \etordef
\eqn\eitc{\mat t{ij}0 = 2g_{ij}\equiv 2\delta_{ij}\quad,}
and any two such frames are related by \eFtm\ with the matrix $M$
orthogonal. Indeed \eFtm\ shows that we have a lot of freedom with
$E_0$; modifying it by adding any linear combination of the $\vec E$
does not change the \SC\ structure. Given a normalized frame we can
thus discard $E_0$ and focus on $\vec E$, regenerating $E_0$ when needed by
$\half [E_+,E_-]$ or some other convenient variant.

Recall that a complex
structure has been given on the manifold and that $\{ E_0, \vec E \}$
are holomorphic. Hence in an arbitrary complex
coordinate system with coordinates given by $w$ and $\lambda^i$, we can
represent $\{E_i\}$ by
\eqn\eSc{E_i={M_i}^j \pa_j + \alpha_i \pa_w\quad,
}
where $\pa_i\equiv\pd{}{\lambda^i}$, $\pa_w\equiv\pd{}w$ and
 ${M_i}^j$ and $\alpha_i$ are holomorphic functions of $w$ and
$\lambda^i$.

We would like to show that we can find a coordinate system in which
$\{E_i\}$  is $G^N$-equivalent to the standard frame $\{D_i\}$. We shall
proceed in four steps,  order by order in the odd
coordinate $\lambda$.

\noindent
{\bf Step 1:} We shall first find a coordinate system in which
\eqn\eEi{
E_i=\pa_i + \ordl\quad.
}
Let $\mat{M}ij\ = \mat mij\ + \ordl$ and $\alpha_i = \alpha_{i0} +
\ordl$ in~\eSc.
We make the following complex coordinate transformation:
\eqn\ectra{
\tilde\la^i = \la^j{{[m\inv]}_j}^i\quad;\quad\tilde w=w\quad.
}
Under coordinate transformation \ectra, we obtain that
$$
E_i = \tilde\pa_i + \alpha_{i0} \tilde\pa_{w} + \ord{\tilde\lambda}\quad.
$$
We can now drop the tildes for convenience. We make another complex coordinate
transformation
\eqn\ectrb{
\tilde\la^i=\la^i\quad;\quad\tilde w= w + \la^r\beta_r\quad,
}
and obtain
$$
E_i = \tilde\pa_i + (\beta_i +\alpha_{i0})\tilde\pa_w+\ordl\quad.
$$
Choosing  $\beta_i=-\alpha_{i0}$, we obtain (after dropping the tildes
again) \eEi.

\noindent
{\bf Step 2:} Restoring $\la$ terms in $E_i$, we have
\eqn\efrb{
E_i= \{ \mat\delta ik + \la^r\mat\mu{ri}k \}\pa_k + \la^ra_{ri}\pa_w
+\ord{\la^2}\quad,
}
where we have introduced two functions $\mat\mu{ri}k$ and
$a_{ri}$.
The normalization conditions \eitc\ are easily seen to imply that
$$a_{(ij)}=\delta_{ij}a_0$$
where $a_0$ is some invertible function. The antisymmetric part of
$a_{ij}$ can now be removed by a coordinate transformation of the form
$$\tilde w=w +\half\lst b_{st}\quad,$$
while the trace bit can be set to one by a further transformation of
the form $\tilde w=\tilde w(w)$ with $\pd{\tilde w}w=a_0\inv$.
We will now use our freedoms to put %
${\mu_{ri}}^k$ into more canonical form.

Again we perform coordinate transformations. Let
\eqn\ectrc{
\tilde\la^i =\la^i + \lfr1{2!}
\la^r\la^s\mat\rho{sr}i\quad;\quad\tilde w =w\quad,
}
where $\mat\rho{sr}i=\mat\rho{[sr]}i$.
In this coordinate system
\eqn\etwoa{
E_i =\{ \mat\delta ik + \tilde\la^r(\mat\rho{ri}k +
\mat\mu{ri}k)\}\tilde\pa_k + \tilde\la_i\pa_{\narrowtilde w}
+\ord{\tilde\la^2}\quad.
}
We can also consider the $G^N$-equivalent frame
$E_i' = (KE)_i$, where
\eqn\etwotm{\mat Kij=\mat \delta ij
+\lambda^r(\mat\alpha{ri}j+\xi_r\mat\delta ij) +\CO(\lambda^2)\quad.}
Here $\alpha_{rij}=\alpha_{r[ij]}$ is a generator of $SO(3,\spec C)$.
Together with \ectrc\ we see that we can shift $\mu$ by
$$\mu\to \mu_{rik} + \rho_{rik} + \alpha_{rik} + \xi_r\delta_{ik}\quad.$$
To begin simplifying this we see we may without loss of generality use
$\rho$ to get $\mu=\mu_{(ri)k}$, symmetric on the first two indices. A
little algebra then shows that with an appropriate choice of further
$\rho,\alpha$ transformations we may take $\mu=\mu_{(rik)}$, and
moreover using $\xi$ we can get $\mat\mu{ij}i=0$.

\noindent {\bf Step 3:} Thus we
have $$E_i=D_i+\la^r\mat\mu{ri}k\pa_k+ \half\lst \epdn tsu(\mat M{ui}k
\pa_k + \Theta _{ui} \pa_w)+ \order3\quad,$$
where again $\mu=\mu_{(rik)}$ and we have introduced the next order,
coefficients $\mat M{ui}k$ and $\Theta _{ui}$.

Using our freedom to choose a convenient $E_0$ we now take
\eqn\eezpn{E_0=\lfr16\sum_i[E_i,E_i] + F^\ell E_\ell\quad,}
where $F^\ell$ is some function of $w,\lambda$ of order $\lambda$.
Imposing \eitc\ to $\order{}$ now shows that $\mu\equiv0$ and
$\Theta _{ui}\propto\delta_{ui}$. But this means that we may remove $\Theta $
altogether by the coordinate transformation
$\tilde w=w + \la^3\beta$, where
$$\la^3\equiv\lfr16 \lst\lambda^u\epdn uts\quad.$$

\noindent {\bf Step 4:} Thus we have
$$E_i=D_i + \half\lst\epdn tsu\mat M{ui}k\pa_k + \la^3(s_i\pa_w +
\mat\sigma i\ell\pa_\ell) $$
where $s_i,\mat\sigma i\ell$ are new sets of coefficients. There
remain the freedom to make coordinate transformations of the form
$\tilde\la^i=\la^i + \la^3K^i$ as well as $SO(3,\spec C)\times\spec
C^\times$ frame rotations. One readily sees that this freedom suffices
to make $\sigma$ traceless symmetric, $M=M_{u(ik)}$, $s_i\equiv0$, and
$\mat M{ik}i=0$.

We now make a convenient choice of $F^\ell$ in \eezpn:
\eqn\eezpnb{E_0=\lfr16\sum_i[E_i,E_i]-\lfr16 (2\lambda^s\epdn siu \mat
M{ui}\ell + \lst\epdn tsi\sigma^{i\ell})E_\ell\quad.}
Then the condition \eitc\ says $\sigma\equiv0$, $M\equiv0$. Thus we have
$$E_i=D_i$$
as was to be shown.

We close this section by remarking that \SC\ integrability should be
related to the conformal flatness of an appropriate \s gravity theory.
Indeed $N=3,4$ \s gravity theories have been constructed using
conformal flatness as a principle~\ref\GHvN{S.J.Gates, Y.Hassoun
and P.van Nieuwenhuizen, Nucl. Phys. {\bf B317} (1989) 302. }. Perhaps
the rather simple idea of \SC\ geometry can shed some light on the
structure of these theories.

\def\SR{semi\-rigid}
\def\sc{super\-conformal}
\def\ct{coordinate transformation}
\def\LI{L^{-1}}
\def\tp{\tilde \theta^+}
\def\tm{\tilde \theta^-}
\def\tt{\tilde \theta^3}

\newsec{Semirigid Integrability}
\subsec{TN=0 Integrability}
We now investigate the local integrability of \SR\ structures. To begin
suppose we have been given a $TN=0$ (or ``almost
semirigid'') structure specified by a frame
$\{E_0, E_+,E_-\}$ obeying \eitc. This is the same information as
in the \SC\ case, but now we do not consider two frames equivalent
unless they are related by \essr, \ie
\eqn\enttg{\pmatrix{E'_0\cr E'_+\cr E'_-}=
\pmatrix{c&\multispan2\hfil$\ \ \cdots$\hfil\cr
0&1&0\cr
0&0&c\cr}\pmatrix{E_0\cr E_+\cr E_-}\quad.}
Thus to integrate the frame we have
a harder job than in sect.~3: find local coordinates such that the
standard frame equals the given one modulo $G^{TN=0}\subset G^{N=2}$
\enttg, not just modulo $G^{N=2}$ \eFtm.

We can again simplify the problem somewhat by noticing that $E_0$ will take
care of itself once we put the $\vec E$ into the desired form.
Accordingly we take $E_0=\half
[E_+, E_-]$, since this choice is still normalized correctly and is
related to the given one by a $G^{TN=0}$ transformation \enttg.

Following the procedure in sect.~2.2, we look for torsion
constraints by taking a standard frame and applying an arbitrary
transformation of the form \enttg: %
$E_+=D_+$, $E_-=cD_-$ where $c$ is a function. By the remark in the
previous paragraph we then take $E_0=\half
[E_+, E_-]$. Computing $\t ABC$ we find that in
addition to \eitc, preserved since we have maintained the
normalization condition, we also preserve various other elements of
$t$, including in particular \eqn\etctga{\t+++=0\quad.}
Thus \etctga\ is necessary for a frame to be integrable.

Now suppose our given frame does satisfy \etctga. By the
result of the previous section we can at least find \SC\ coordinates,
\ie\ coordinates $\bz=(z,\theta^\pm)$
 such that $\{E_0,\vec E\,\}$ is $G^{N=2}$-equivalent
to $\{\pa_z,\vec D\,\}$. In particular
\eqn\etgfir{ D_i= \mat Nij E_j\ \,\quad N = \pmatrix{ab^{-1}&0 \cr
                                0&ab \cr } \in SO(2,{\spec C})}
for some invertible functions $a,b$.
We would now like to find another set of coordinates $\tilde\bz(\bz)$
with
\eqn\etgsec{ \tilde D_i= \mat{(K\inv)}ij E_j\ \,\quad K = \pmatrix{1&0 \cr
                                0&c \cr } \quad.}
Since $\bz$ is not a good \SR\ coordinate, $\tilde\bz(\bz)$ is not a
\SR\ transformation. However \etgfir--\etgsec\ say $\bz$ and
$\tilde\bz$ are at least \SC\ coordinates and so
$\tilde\bz(\bz)$ will be a \SC\
transformation. But we know how $\vec D$ transform under the latter
(eqn.~\eMtm).

Putting it all together, given $a,b$ in \etgfir\ we need to choose $\tilde \bz$
and $c$ in \etgsec\ such that
\eqn\etgsolv{\pmatrix{1&\cr &c\inv\cr}\pmatrix{b/a&\cr &(ab)\inv\cr}
\pmatrix{D_+\cr D_-\cr} = \pmatrix{D_+\tilde\theta^+&\cr
 &D_-\tilde\theta^-\cr}\inv\pmatrix{D_+\cr D_-\cr}\quad.}
In other words, while we can always adjust $c$ to satisfy the second
equation, we do need to find a \SC\
transformation for which $D_+\tilde\theta^+=a/b$. As
expected we see that in general there is no solution. Imposing
\etctga, however, tells  that $D_+(a/b)=0$, which
 ensures that an appropriate function $\tilde\theta^+$
exists. To see that there is a $\tilde\bz$ with this $\tilde\theta^+$,
we need to inspect the most
general $N=2$ \sc\ \ct:  %
\eqn\entsc{\eqalign{ {\tilde z} &= f + \theta^+ t \psi
+ \theta^- s \tau  +\theta^+ \theta^- \pa_z (\tau \psi) \cr
{\tilde\theta}^+&=\tau +\theta^+ t +\theta^+ \theta^- \pa_z \tau \cr
{\tilde\theta}^-&=
\psi +\theta^- s -\theta^+ \theta^- \pa_z \psi \quad ,}}
where $\pa_z f =t s - \tau \pa_z \psi - \psi \pa_z \tau $.
Thus, we have
\eqn\edptp{ D_+ {\tilde \theta}^+ = t + 2\theta^-\pa_z \tau -
\theta^+ \theta^- \pa_z t }
and we can choose $t,\tau$ to match this to any chiral superfield $a/b$.

\subsec{TN=1 Integrability}

In this subsection, we start with an $N=3$ $SRS$ endowed with
the $TN=1$  structure given by a frame $\{E_0,E_\pm,E_3\}$
normalized per \eitc, \eoffbasis. As in the previous subsection we
may discard $E_0$ and replace it by $E_0=\half[E_3,E_3]$ without
changing the \SR\ structure.

Proceeding as before we get the torsion constraints by
acting on  the standard frame $\vec D$ with\foot{Recall that in this
basis the metric $g_{ij}$ is antidiagonal.}
\eqn\eGmat{K = \pmatrix{1&x&{-{x^2\over 2}} \cr
             0&a&{-ax} \cr  0&0&a^2  \cr }  }
where $a$ is invertible.
Examining the commutators of $E_i=\mat KijD_j$ we find that in addition
to \eitc\ we have (among other things)
\eqn\enthrtc{\t ij+=0\ ,\quad \t--3=0\quad.}
We will show that these necessary conditions are sufficient for integrability.

Suppose then that we have a local frame $\vec E$ for \CCE \ obeying
\eitc\ and \enthrtc\ once we set $E_0=\half[E_3,E_3]$. Once again
we can use the result in sect.~3 to choose \SC\ coordinates, so that
$E_i = {(N^{-1})_i}^j D_j$ where $N$ belongs to
$SO(3,{\spec C})\times {\spec C}^\times$.
Semirigid integrability means
that there exists a \sc\ \ct\ $\tilde\bz(\bz)$
such that the given frame $E_i$ is $G^{TN=1}$-equivalent to
the standard frame ${\tilde D}_i$:
$$(K^{-1}E)_i = {\tilde D}_i $$
with $K$ some matrix function of the form \eGmat. Analogous to
\etgsolv\ this requires us to solve
\eqn\eFNG{F=NK\quad {\rm where}\quad{F_j}^k=D_j\tilde\theta^k\quad.}
In this equation  we are seeking suitable
$\tilde\bz(\bz)$ and $K$ given $N$. Once again this is in general
impossible until we impose the constraints \eitc, \enthrtc\ on $N$.

We will subdivide our task by writing $K=K_1K_2$ and choosing $K_1$ to put
$\hat N=NK_1$ into the form of a lower triangular matrix $L$ (when
${N_{+}}^{+}$ is invertible) or an ``upper'' triangular matrix $U$
(when ${N_{+}}^{+}$ is not invertible; see below) with unit determinant.
This puts our problem \eFNG\ into standard form: $F=L K_2$ or $=UK_2$.
We will in the following concentrate on the case when
${N_{+}}^{+}$ is invertible, prove the integrability theorem and
then comment on the other case.

We organize the proof into four steps.
First, we will show that $NK_1$ can be put into the form of
$L$.  Then we impose the \SR\ essential torsion constraints
 on $L$ (recall the constraints are $G^{TN=1}$-invariant).
With this done, we will substitute $L$ into \eFNG, and
solve for the $\tp$ component of the \sc\ \ct\ in terms of the
unconstrained superfield components of
$L$ just as in sect.~4.1. The rest of the
components, $\tt$ and $\tm$,
can always be made to satisfy \eFNG\ by choosing $K_2$
appropriately. %
Finally, we will show  by construction that %
there really does exist a \sc\ \ct\ with the required $\tp$.

The torsion constraint \eitc\
implies that $N$ belongs to $SO(3,{\spec C})\times {\spec C}^\times$, meaning
$NgN^t \propto g$.
In particular we have %
\eqn\esonpm{{N_+}^- = -{({N_+}^3)^2 \over {2 {N_+}^+}}
 \quad {\textstyle and} \quad
 {N_3}^- = {{N_3}^+ \over 2}
( {{N_+}^3 \over {N_+}^+ } )^2
 -{ {{N_+}^3 {N_3}^3} \over {N_+}^+ } \quad. }
${({NK_1})_{+}}^{3}$ is set to zero by choosing $K_1$ in \eGmat\ with
\eqn\exga{ x_1=-a_1 { {N_+}^3 \over {N_+}^+} .}
Substituting \esonpm\ and \exga\ into the
expressions ${(NK_1)_{+}}^{-}$ and ${(NK_1)_{3}}^{-}$, they too vanish.
Furthermore, $a_1$ is chosen so that $NK_1$ has unit determinant,
that is, $a_1= (\det N)^{-{1\over 3}}$.  Thus, $NK_1$ by construction
is a lower triangular matrix given by
\eqn\ematL{L=\pmatrix{b&0&0 \cr y&1&0 \cr
      {-{y^2 \over {2b}}}&{-{y \over b}}&{1\over b} \cr}\quad.}

We can now let $\tilde E_i=(K_1\inv E)_i=(L\inv D)_i$. While we have
used our $G^{TN=1}$ freedom to put $L$ into the standard form \ematL, still
further restrictions come when we impose the
 torsion constraints \enthrtc.
Since these
torsion constraints are by construction $G^{TN=1}$-invariant, we
can impose them on $\tilde E_i$. %
The constraints give respectively
\eqn\etmmt{ {(\LI)_-}^k ~D_k {(\LI)_-}^l {L_l}^3 =0 \quad{\rm and}}
\eqn\etijp{ \{ {(\LI)_{(i}}^k ~D_k {(\LI)_{j)}}^l
           - \optional{2}g_{ij} {(\LI)_3}^k~ D_k {(\LI)_3}^l \}~ {L_l}^+ = 0.}
Substituting \ematL\ into \etmmt\ and \etijp, we obtain the following
four constraints on the two independent matrix elements
$b$ and $y$ of the matrix $L$:
\eqn\eDpb{ D_+ b = 0,}
\eqn\eDpy{ D_+y + D_3 b= 0, }
\eqn\ebDmb{ bD_-b -2bD_3y - yD_+y =0, \quad {\rm and}}
\eqn\eysDpy{ {y^2 \over 2}D_+y + byD_3y - b^2 D_-y=0.}
In appendix A, we show that under this set of torsion constraints
we obtain a unique odd superfield $\Omega$ satisfying
\eqn\edom{b=D_+ \Omega\ ,\quad y=D_3 \Omega\ ,\quad {\rm and\ }
g^{ij}(D_i\Omega)(D_j\Omega)=0\quad.}

We are now ready to show that there
exists a \sc\ \ct\ and suitable $K_2$ for which $F=LK_2$.
That is,
\eqn\eLGF{\pmatrix{D_+\tp & D_+\tt & D_+\tm \cr
                  D_3\tp & D_3\tt & D_3\tm \cr
                  D_-\tp & D_-\tt & D_-\tm  }
        = \pmatrix{b & bx_2 & {-bx_2^2\over 2} \cr
                   y & x_2y+a_2 & -x_2(a_2 + {{x_2y}\over 2}) \cr
   -{y^2\over 2b} & -{y\over b}(a_2 + {{x_2y}\over 2}) &
                            {1\over b}(a_2 + {{x_2y}\over 2}) },}
where
$a_2$ and $x_2$ are the independent elements
of $K_2$.
Taking the determinant of both sides of \eLGF,
we see that we have to choose
\eqn\eat{ a_2 = \det(F)^{1\over 3}.}
As for $x_2$, we will choose it so that $bx_2 = D_+\tt $.
Eqn.~\edom\ then shows that
the first column of equations \eLGF\ are satisfied when we
identify $\tp$ as $\Omega$.
One can show that
the remaining five components of the matrix equation \eLGF\ are then
satisfied by the use of the \sc\ conditions \escc. These turn into
 two sets of readily applicable relations
\eqn\eFson{FgF^t = g{(\det F)}^{2\over 3}}
and the set of equations where we replace $F$ by $F^{-1}$,
since $F^{-1}$ is also a \sc\ transformation.

Finally, the question is if there exists an  $N=3$
\sc\ \ct\ with $\tp$ given by the function $\Omega$.  The answer is
yes; details are given in appendix B.  The
point is that from the \sc\ conditions ${\tilde z}$ can be expressed
in terms of the components of the transformation of
$\tilde \theta^i$, $i=+,3,-$.
The only requirement left for the \ct\ to be
\sc\ is that the $\tilde \theta^i$ satisfy the \sc\
conditions among themselves.
In  appendix B, we have expanded ${\tilde z}$
and $\tilde \theta^i$ in components.  There are four even and
four odd components in each of the superfields.
We set out with $\tp$ given, namely $\Omega$,
and there are sixteen degrees of freedom in the components of $\tt$ and
$\tm$ to choose to satisfy the internal \sc\ conditions.
The \sc\ conditions among $\tilde \theta^i$
are linear in the components of $\tt$ and $\tm$ and there are
sixteen such equations.
We have shown in the appendix that indeed a solution exists.
If all the even components of $\tp$ are invertible,
then we use all sixteen degrees of
freedom to solve the sixteen equations. If one or more even components
of ${\tilde \theta^+}$ are noninvertible, then the linear matrix
equations become singular and it implies that there are more
variables than equations. Thus,
there exists a family of solutions.

When ${N_+}^+$ is not invertible, from the fact that $N$ belongs to
$SO(3,{\spec C}) \times C^\times$, we immediately obtain that
${N_-}^+$, ${N_+}^-$ , and ${N_3}^3$ are invertible.  Since
${N_-}^+$ is invertible, we can choose elements in $K_1$ so that
$(NK_1)$ takes the form %
\eqn\eu{U=NK_1 = \pmatrix{ -{y^2\over {2b}}& -{y\over b}&{1\over b} \cr
                           y&1&0 \cr b&0&0 \cr}.}
All the essential torsion constraints are the same as before with
the roles of $D_+$ and $D_-$ interchanged.  We again wish to find a
\sc\ \ct\ $F$ so that \eFNG\ is satisfied.  We then have
$b=D_-{\tilde \theta^+}$ and $y=D_3{\tilde \theta^+}$.  The rest of the
proof is analogous to the previous case  with the roles
of the superfield components switched between the untilded $+$ and $-$
components and a sign change for the tilde components along with
interchanging the $+$ and $-$ components
(\eg\ $s_- \rightarrow s_+$ and ${\tilde \psi_-} \rightarrow
-{\tilde \psi_+})$.

\newsec{Moduli space of \SR\ surfaces}

There exists a natural projection from the
moduli space of $TN$-\SR\ surfaces to that of $N$-$SRS$ \semirig\topsug.
We will show that this is the case for $TN=0,1$. This can be easily
extended for the case of arbitrary $N$. As
explained earlier, an $N$-$SRS$ is obtained
by patching together pieces of ${\spec C^{1|N}}$ by means of
$N$-superconformal transformations:
\eqn\emodu{
{\bz _\alpha} = f_{\alpha\beta}({\bz _\beta}; \vec m, \vec\zeta\,)
}
where ${\bz }=(z,\vec\theta\,)$ and $\vec m$ ($\vec\zeta\,$) are the even
 (odd) moduli.
Following~\semirig\dil\ we obtain {\it augmented}
$N$-superconformal transformations by introducing a new global odd
variable $\theta^+$ and promoting all the functions given above to be
arbitrary functions of $\theta^+$ in addition to ${\bz }$.
Now an {\it augmented}
$N$- superconformal surface is obtained by patching together pieces of
${\spec C^{1|N+1}}$ by means of the augmented superconformal
transformations. An augmented $N$-superconformal surface still has a
distinguished distribution $\bar\CE$ of dimension $0|N$
spanned by $\vec D$. This is
seen by checking that under augmented superconformal transformations,
$\bar\CE$ is preserved.

The group of augmented $N$-superconformal transformations
is isomorphic to  $TN$-\SR\ transformations. This has been proved
for the cases of $TN=0$%
{}~\semirig\ and
for $TN=1,2$ \topsug.
Since we may represent any $SSRS$ by a collection of semirigid
patching
functions, we can apply this isomorphism to obtain an augmented $SRS$ and
vice versa.\foot{Note however that as a complex \s manifold the
$TN$-surface is of dimension $1|N+2$ while the corresponding augmented
$N$-$SRS$ is of dimension $1|N+1$. The missing $\theta^-$ carries no
information, though it was crucial to get superfield formulas in \semirig.}
\ This isomorphism implies that the moduli spaces of $TN$-$SSRS$
and augmented $N$-$SRS$ are identical.
Hence, it suffices to study the moduli space of augmented
$SRS$.

The moduli of the augmented superconformal surfaces are obtained by
replacing the moduli of the superconformal surfaces by functions of
$\theta^+$ in \emodu, that is
\eqn\eaugmod{
\eqalign{
m^a &\mapsto \tilde m^a + \theta^+ \hat m^a\cr
\zeta^\mu &\mapsto\tilde \zeta^\mu + \theta^+\hat\zeta^\mu
}}
where we have introduced extra odd(even) moduli,
$\hat m^a$ ($\hat\zeta^\mu$) and placed tildes on $\tilde m^a$
($\tilde\zeta^a$)
to avoid confusion with the $m^a$ ($\zeta^a$) on the original space.
Hence, given any family of $N$-$SRS$, we obtain a family of
augmented $N$-$SRS$ with twice as many parameters. We lack global information
regarding the moduli space of augmented superconformal surfaces. For
example, we do not know if any of the new even coordinates $\hat\zeta^\mu$
are periodic. But we can easily argue that infinitesimally, \eaugmod\
spans the full tangent to the moduli space when we vary $\tilde m,~\hat m,~
\tilde\zeta,~\hat\zeta$. First, we note that deformations of any
 augmented $SRS$
involve small changes in the patching maps. These are generated by
vector fields $V_{\alpha\beta}$ on $U_\alpha\cap U_\beta$
with no $\theta^+$ component (this follows from
the global nature of $\theta^+$). Expanding $V_{\alpha\beta}$ in a power series
in
$\theta^+$, we get two identical copies of the deformation space of $N$-$SRS$,
with opposite parity.
Furthermore, given
$V^A(z,\vec\theta, \theta^+) = v^A(z,\vec\theta\,)
 + \theta^+\nu^A(z,\vec\theta\,)$ with $V^{\theta^+}=0$,
the vector field $\{v^A_{\alpha\beta}\}$ generates infinitesimal deformations
in the moduli $\tilde m$ and $\tilde \zeta$ and the vector field
$\{ \nu^A_{\alpha\beta} \}$ generates infinitesimal deformations in
$\hat m$ and $\hat\zeta$ from \eaugmod.

A projection down to the moduli space of $N$-$SRS$
corresponds to forgetting the new moduli introduced, that
is, given a point with coordinates $(\tilde m^a, \hat{ m}^a, \tilde\zeta^\mu,
\hat{\zeta}^\mu)$ in the
augmented moduli space, we project down to the point with coordinates
($ m^a=\tilde m^a, \zeta^a=\tilde \zeta^a)$ in the moduli space of
$SRS$. We would like to show that  the projection is natural.

Let us now discuss projections in general. We wish to define a map $\Pi$ from a
space \CCMH\ to \CCM.
Let ($\tilde  x^a,~ \hat x  ^a$) be a set of coordinates near
$\tilde P$ on $\hat{\cal M}$ and
${x}^a$ be coordinates near
${P}$ on $\cal M$. We can define a projection $\Pi$
by taking $ x^a(P)=\tilde{x}^a (\tilde P)$ or in other words
$$
\Pi^*({x}^a) =\tilde  x^a
$$
which we refer to as the ``forgetful'' map.
Unfortunately, the definition of $\Pi$ depends on the choice of
coordinates.
Let $(\tilde y^a=\tilde F^a(\tilde  x^b,{\hat x  ^b}\,),~
{\hat y^a}=\hat F^a( \tilde x^b,{\hat x^b  }\,))$ be another set of coordinates
near $\tilde P$.
Also, let $ {y^a}= F^a(\vec {{x}})$ be a new coordinate near
${P}$. The new coordinates will define the same map $\Pi$ as the old ones
only if
\eqn\ectproj{\tilde y^a\equiv\tilde F^a(\tilde x^b,\hat x^b \,)= F^a( \tilde
x^b)\quad.}
Of course arbitrary coordinates for
\CCMH\ will not be related to $(\tilde x^a,{\hat x^a  }\,)$ by \ectproj. But
if \CCMH\ has some natural class of coordinates all related by
\ectproj\ then we do obtain a global projection $\Pi$. We will now see
that \SR\ moduli space does have such a natural class of coordinates.

Begin with the case of $TN=0$ following the discussion in
\semirig. A Riemann
surface is obtained by patching together pieces of ${\spec C^1}$ by
means of the transition function
$$
z_\alpha =  f_{\alpha\beta}(z_\beta, {m}^a)\quad,
$$
where ${m}^a$ are complex
coordinates on the moduli space of
complex dimension $(3g-3)$. We now obtain a class
of augmented Riemann surfaces parametrized by $(\tilde m^a,{\hat m}^a)$
using the augmented transition functions
\eqn\epata{ \eqalign{
     z_\alpha &=  f_{\alpha\beta}(z_\beta, \tilde m^a+\theta^+_\beta\hat m
^a)\quad,\cr
                &=  f_{\alpha\beta}(z_\beta,\tilde m^a) + \theta_\beta^+
                     \pa_af_{\alpha\beta}(z_\beta,\tilde m^a)\hat m ^a\quad;\cr
\theta_\alpha^+ &=  \theta_\beta^+\quad,}
}
where a point in
the moduli space of augmented Riemann surfaces has coordinates
($\tilde m^a$, $\hat m ^a$). Coordinates obtained in this way are not the most
general ones, and indeed we will now show that they are all related by
the special class of maps \ectproj.

Let $ n^a( m^a)$ be a new set of coordinates
 on the moduli space of ordinary Riemann surfaces. We
obtain the patching function parametrized by $ n^a$
by means of the following identification:
\eqn\ecompa{
\check f_{\alpha\beta}(z_\beta, \vec n)\equiv f_{\alpha\beta}(z_\beta,
\vec m({\vec n}))\quad.
}
The corresponding family of augmented Riemann surfaces is again given
by the rule \eaugmod:
\eqn\epatb{ \eqalign{
       z_\alpha &=  \check f_{\alpha\beta}(z_\beta,
                   \tilde n^a+\theta^+\hat n^a)\quad\cr
                &=  \check f_{\alpha\beta}(z_\beta, \tilde n^a) +
\theta_\beta^+
                     \pa_a\check f_{\alpha\beta}(z_\beta,\tilde n^b)\hat n^a
\cr
         \theta_\alpha^+ &= \theta_\beta^+\quad.}
}
Comparing \epata\ and \epatb\ using \ecompa\ shows that the two sets of
coordinates
on the moduli space of $TN=0$ surfaces are related by the transition function
$$\eqalign{
     \tilde n^a &=\tilde  n^a( \tilde m^b)\quad,\cr
    \hat n^a &=  \left.{{\pa n^a  } \over {\pa m^b}}\right|_{m=\tilde m} \hat m
^b\quad,}
$$
which is not only of the form \ectproj\ but in fact split. Hence in
particular the projection from augmented $N=0$ surfaces to ordinary
ones is natural, and as we have already seen that this gives the
desired projection from $TN=0$ surfaces to $N=0$.

For the case of $TN=1$, the situation is similar.
Let ($\tilde m^a,~\hat m^a,~\tilde \zeta^\mu,~\hat\zeta^\mu$) be the
coordinates of
a point in the moduli space of augmented $N=1$ $SRS$ and $(\tilde n^a,\hat n^a,
\tilde \phi^\nu,\hat\phi^\nu)$ %
be the coordinates of the same
point on another patch. Following similar arguments as for $TN=0$,
we obtain
$$\eqalign{
      \tilde n ^a &=  \tilde n ^a(\tilde  m^b,\tilde \zeta^\nu)\quad,\cr
    \hat n ^a &=
    {{\pa  n ^a  } \over {\pa m^b}} \hat m^b +
    {{\pa  n ^a  } \over {\pa \zeta^\nu}} \hat\zeta^\nu\quad,\cr
     \tilde \phi^\mu &= \tilde \phi^\mu(\tilde  m^b,\tilde \zeta^\nu)\quad,\cr
    \hat\phi^\mu &=
    {{\pa \phi^\mu  } \over {\pa m^b}} \hat m^b +
    {{\pa \phi^\mu  } \over {\pa \zeta^\nu}} \hat\zeta^\nu\quad,}
$$
which is again of the form \ectproj\ and hence the ``forgetful'' map
is again natural.
This can be seen to hold for the case of arbitrary $TN$ since the
only property which makes the transition function split is the
global nature of $\theta^+$. Thus, there exists a natural projection
from the moduli space of $TN$-$SSRS$ to the moduli space of $N$-$SRS$.
The significance of this result is that~\semirig\ it means we can use
string-theory methods to get a measure on the big space, then
integrate it over the fibers of this projection to get a measure on
the smaller space, namely the moduli space of $N$-\SC\ surfaces, which
is where the observables of \TG\ should live.

\newsec{Conclusion}
In this paper, we have provided an intrinsic definition of $N$-$SRS$ and
$TN$-$SSRS$ which appeared naturally in (super)gravity and
topological (super)gravity respectively.  The intrinsic
definitions are given in the context of $G$-structures.  It is straight%
forward to define
superconformal or semirigid $G$-structure from the coordinates given in
the patch definition of $SRS$ or $SSRS$.  Much of our analysis was devoted
 to showing how one can recover the patch definition given a
 $G$-structure on a manifold. That is, we first obtained the necessary
torsion constraints where needed and showed that the
almost $G$-structure is integrable under such conditions. %

Moreover, we have shown that there exists a natural projection from the
moduli space of $TN$-$SSRS$ to that of $N$-$SRS$.  Since a field
theoretical realization of topological TN-gravity can yield an
integration density on the moduli space of $TN$-$SSRS$, the natural
projection allows us to integrate along the fibers of the projection and
obtain an integration density on the moduli space of $N$-$SRS$.  If
there are non-trivial observables, then the field theory provides for
us cohomology classes on the moduli space, thus probing its topology.
This procedure has been used
for the case $TN=0$ in~\dil\punc; it would be interesting to see
what topologies one can probe for $TN \ge 1$ cases.
\vskip.8truein {We would like to thank J. Distler, B. Ovrut and S-J. Rey
for useful conversations.
 This work was supported in part by NSF grant PHY88-57200,
DOE contract DOE-AC02-76-ERO-3071 and by the A.~P.~Sloan
Foundation.}

\appendix{A}{Solving the torsion constraints}
We will show that under the constraints \eDpb\ to
\eysDpy, we can find a unique $\Omega$  to represent $b$ and $y$ by
$D_+\Omega$ and $D_3\Omega$ respectively
and satisfying the constraint
\eqn\ecO{D_+\Omega~D_-\Omega = -{1\over 2} (D_3\Omega)^2.}
Anticipating Appendix B we note that if $\Omega=\tilde\theta^+$ then
\ecO\ is one of the \sc\ conditions involving only $\tp$
in \eFson\ when $F^{-1}$ is used.

We will now go through the constraints \eDpb\ to \eysDpy\ and show
how we get $\Omega$.  Eqn.~\eDpb\ implies that $b=D_+\Omega$ where
$\Omega$ is an odd superfield.  To see that is possible, one way is
to expand both $b$ and $\Omega$ in components $\theta^i$
and constrain $b$ by \eDpb.  Then it is straightforward that
equating $b$ and $D_+\Omega$ turns to algebraic equations between their
components, thus solving for the components of $\Omega$ in terms of
that of $b$.
However there is a residual freedom
$\Omega \rightarrow \Omega'=\Omega + \omega$ where $D_+\omega = 0$
leaves $b=D_+ \Omega$ invariant.
We will make use of this degree of freedom
to make $y=D_3\Omega$.  Substituting $b=D_+\Omega$ into \eDpy, it implies
that $y=D_3\Omega + B$, where $D_+B=0$.  Here we will use the
freedom in choosing $\Omega'$ to cancel $B$, that is, $D_3\omega=-B$.
This is possible because both $\omega$ and $B$ are annihilated by
$D_+$, and in components, it means solving two algebraic equations
and two first order linear differential equations in the
components of $\omega$ in terms of that of $B$.
There is still a little freedom left in $\Omega'$,
namely $\Omega''= \Omega' + \phi$, where $D_+\phi=D_3\phi =0$.
In components, this means $\phi=\theta^- \phi_-$, and $\phi_-$
is a constant.  Again, this constant will be used later on.
Dropping the primes,
we now have $b=D_+\Omega$ and $y=D_3\Omega$; substituting both into
\ebDmb\ and \eysDpy, we obtain
\eqn\ebDmbO{D_3\Omega~D_+ D_3\Omega = - D_+\Omega~D_+ D_- \Omega}
and
\eqn\eysDpyO{ -{1\over 2} (D_3\Omega)^2~D_3 D_+\Omega
      +D_+\Omega~D_3\Omega~\partial_z \Omega
      + (D_+\Omega)^2~D_3 D_-\Omega=0  .}
Eliminating
$D_3\Omega~D_+ D_3\Omega$ in \eysDpyO\ by \ebDmbO, we get
\eqn\ecombO{ -{1\over 2}D_3\Omega~D_-D_+\Omega + D_+\Omega~D_- D_3\Omega
            =0.}
Equations \ebDmbO, \eysDpyO\ and \ecombO\ can be rewritten as
\eqn\emagic{D_i \lbrack D_-\Omega
     + {1\over 2}{{(D_3\Omega)^2}\over {D_+\Omega}} \rbrack =0 ,}
where $i=+,3,-$ respectively.
This implies that whatever is inside the square bracket can at
most be some arbitrary constant.  This constant can be cancelled
by the remaining free constant $\phi$. By construction, $D_3$ and
$D_+$ annihilate $\phi$, and $D_-\phi=\phi_-$.  Thus, $\phi_-$
will be chosen to cancel the arbitrary constant, and we are left
with \ecO.

\appendix{B}{N=3 \sc\ \ct}
In this appendix, we will give the conditions for the $N=3$ \sc\ \ct.
We will then show that there exists an $N=3$ \sc\ \ct\ when
$\tilde \theta^+$ is given subject to \ecO.  This is needed in the proof of
$N=3$
semirigid integrability.

We expand the \sc\ transformation in components.  Let
\eqn\ezcp{ {\tilde z} = f
              + \theta^+ \phi_+ + \theta^3 \phi + \theta^- \phi_-
      + \theta^3 \theta^+{\tilde f_+} + \theta^+ \theta^-{\tilde f}
      + \theta^- \theta^3 {\tilde f_-}
      + \theta^+ \theta^- \theta^3 {\tilde \phi} }
 and
\eqn\ethetacp { {\tilde \theta^i} = \lambda^i  + {{(m^t)}^i}_j \theta^j
+{{(\Gamma^t)}^i}_j g^{jk}\half \epsilon_{k\ell m} \theta^\ell  \theta^m
+{\tilde \ell ^i } \theta^+ \theta^- \theta^3, }
where $i=+,3,-$ and
\eqn\em { {m_i}^j = \pmatrix{
   t_+ & n_+ & s_+ \cr
   t & n & s \cr
   t_- & n_- & s_- \cr}\quad,\quad
 {\Gamma_i}^j = \pmatrix{
   {\tilde \tau_+} & {\tilde \nu_+} & {\tilde \psi_+}  \cr
   {\tilde \tau} & {\tilde \nu} & {\tilde \psi}     \cr
   {\tilde \tau_-} & {\tilde \nu_-} & {\tilde \psi_-}  \cr} ,}
\eqn\elleq{ \lambda^i= \pmatrix{\tau \cr \nu \cr \psi \cr} \quad,\quad
        {\tilde \ell}^i = \pmatrix{{\tilde t}\cr {\tilde n} \cr
                                   {\tilde s} \cr }\quad, }
the metric $g=\pmatrix{0&0&1 \cr 0&1&0 \cr 1&0&0 \cr}$, and
$\epsilon_{-3+} = 1$.

The superconformal conditions can be compactly written as
\eqn\esca{ mgm^t = g{(mgm^t)}_{33} \quad,}
\eqn\escb{ {(mg\Gamma^t)}_{ij} = g_{ij} {(mg\Gamma^t)}_{33}
     + \epsilon_{ijk}{(g^{-1}mg\pa_z \lambda)}^k ,}
\eqn\escc{ {(mg{\tilde \ell})}_i = {(\Gamma g \partial_z \lambda)}_i
 - {1\over 4}\epsilon_{ijk} {\lbrack 2\Gamma g \Gamma^t  +
 {(\pa_z m)} g m^t - m g {\pa_z m} ^t \rbrack}_{jk} \quad,}
\eqn\escd{ \pa_z f = {(mgm^t)}_{33} + {(\pa_z \lambda)}^t g \lambda ,}
\eqn\esce{ {(\phi)}_i = {(mg\lambda)}_i \quad,}
\eqn\escf{ {(\tilde f)}_i = {(\Gamma g \lambda)}_i
 \quad, \quad {\rm and} }
\eqn\escg{ {\tilde \phi} = {\tilde \ell}^t g \lambda -
{(mg\Gamma^t)}_{33} \quad.}
There are two things that one notices from \esca\ to \escg.
One is that $m$ belongs to $SO(3,\spec C) \times \spec C^\times$, thus only
four
matrix elements are independent.  The rest can be expressed in terms
of the four independent variables.  The other observation is that
the components of $\tilde z$ of the transformation are
expressed in terms of the components of $\tilde \theta^i$ given by
\escd\ to \escg.  Thus \esca\ to \escc\ are  internal \sc\ conditions
that have to be satisfied by $\tilde \theta^i$.

Our problem is that we are given the components of $\tilde \theta^+$,
and we wish to see that there exists a \sc\ \ct\ with this $\tilde \theta^+$
by choosing the
components of $\tilde \theta^{3,-}$ to satisfy the internal \sc\
conditions.  Let us work with the case when ${N_+}^+$ is
invertible. This implies that $b=D_+ {\tilde \theta^+}$ is also
invertible and hence so is $t_+$. From the lowest component of \ecO,
when $\Omega$ is identified with $\tilde \theta^+$, we have
$t^2=-2t_+t_-$.  Thus even though $\tilde \theta^+$ is handed to us,
we know that $t$ is not independent of $t_+$ and $t_-$.  We will take
$t_+$ and $t_-$ as two independent elements of $m$. There are two left,
and we will choose them to be $s_-$ and $n$.  For now, the only
constraint we put on $s_-$ and $n$ is that they are invertible.
This gives $m$ a invertible determinant.  The rest of the five entries
of $m$ are expressed in terms of $t_+$, $t_-$, $s_-$ and $n$ by \esca.
Since $\tilde \theta^+$ is given to us, we now have, in addition
to $s_-$ and $n$, the rest of the
six elements of $\Gamma$, the lowest and highest components of
$\tilde \theta^3$ and $\tilde \theta^-$ to choose to satisfy
the eight conditions in \escb\ and three in \escc.
Since $t_+$, $s_-$ and $n$ are invertible, we invert them in \escb\
to solve for
$\pa_z \nu$, $\pa_z \psi$, $\tilde \psi_+$,
$\tilde \nu_{+,-}$ and $\tilde \nu$,
thus satisfying six of the eight conditions of \escb.  The two variables
$\tilde \psi_-$ and $\tilde \psi$ have coefficients $t_-$.  $t_-$ is
given to us and it may vanish.
If it does not, then we can invert it and choose
$\tilde \psi_-$ and $\tilde \psi$ to satisfy the last two conditions.
If $t_-$ vanishes then by \esca\ and by \ecO, we conclude that
$n_-^2 = -2s_-t_-=0$,
$t_-=t=\tilde t=\tilde \tau_-=0$ and $\tilde \tau={\pa_z \tau}=
k\tilde \tau_+$, where $k$ is some even function.
Under these circumstances, the two conditions become vacuous.
Similarly, we invert $t_+$ and $n$ in \escc\ to solve for the highest
components of $\tilde \theta^3$ and $\tilde \theta^-$, $\tilde n$ and
$\tilde s$ respectively, thus leaving one condition to be satisfied.
The problem is that we cannot choose $\tilde t$ to satisfy this equation,
but one can see that if $t_-$ is invertible, then we can invert $n_-$ and
choose $\pa_z \tilde n$ to satisfy this condition.  If $t_-$ vanishes,
then this condition becomes vacuous.  Thus, all \sc\ conditions can
be satisfied given $\tilde \theta^+$ and we are able to complete
the rest of the \sc\ \ct.

\listrefs
\bye